
%
%






\newcount\refnumber
\newcount\temp
\newcount\test
\newcount\tempone
\newcount\temptwo
\newcount\tempthr
\newcount\tempfor
\newcount\tempfiv
\newcount\testone
\newcount\testtwo
\newcount\testthr
\newcount\testfor
\newcount\testfiv
\newcount\itemnumber
\newcount\totalnumber
\refnumber=0
\itemnumber=0
\def\initreference#1{\totalnumber=#1
                 \advance \totalnumber by 1
                 \loop \advance \itemnumber by 1
                       \ifnum\itemnumber<\totalnumber
                        \temp=100 \advance\temp by \itemnumber
                        \count\temp=0 \repeat}

\def\ref#1{\temp=100 \advance\temp by #1
   \ifnum\count\temp=0
    \advance\refnumber by 1  \count\temp=\refnumber \fi
     $^{\the\count\temp}$}

\def\reftext#1{\temp=100 \advance\temp by #1
   \ifnum\count\temp=0
    \advance\refnumber by 1  \count\temp=\refnumber \fi
   \ \the\count\temp}

\def\reftwo#1#2{\tempone=100 \advance\tempone by #1
   \ifnum\count\tempone=0
   \advance\refnumber by 1  \count\tempone=\refnumber \fi
   \temptwo=100 \advance\temptwo by #2
   \ifnum\count\temptwo=0
   \advance\refnumber by 1  \count\temptwo=\refnumber \fi
 \testone=\count\tempone \testtwo=\count\temptwo
 \sorttwo\testone\testtwo
     \ [\the\testone,\the\testtwo]}       

\def\refthree#1#2#3{\tempone=100 \advance\tempone by #1
   \ifnum\count\tempone=0
    \advance\refnumber by 1  \count\tempone=\refnumber \fi
    \temptwo=100 \advance\temptwo by #2
   \ifnum\count\temptwo=0
    \advance\refnumber by 1  \count\temptwo=\refnumber \fi
    \tempthr=100 \advance\tempthr by #3
   \ifnum\count\tempthr=0
    \advance\refnumber by 1  \count\tempthr=\refnumber \fi
 \testone=\count\tempone \testtwo=\count\temptwo \testthr=\count\tempthr
 \sortthree\testone\testtwo\testthr
   \test=\testthr  \advance\test by -2
 \ifnum\test=\testone    \test=\testtwo  \advance\test by -1
    \ifnum\test=\testone   
    \ [\the\testone--\the\testthr]\fi \advance\temptwo by 1
  \else
     \ [\the\testone,\the\testtwo,\the\testthr]    
 \fi}

\def\reffour#1#2#3#4{\tempone=100 \advance\tempone by #1
   \ifnum\count\tempone=0
    \advance\refnumber by 1  \count\tempone=\refnumber \fi
    \temptwo=100 \advance\temptwo by #2
   \ifnum\count\temptwo=0
    \advance\refnumber by 1  \count\temptwo=\refnumber \fi
     \tempthr=100 \advance\tempthr by #3
   \ifnum\count\tempthr=0
    \advance\refnumber by 1  \count\tempthr=\refnumber \fi
    \tempfor=100 \advance\tempfor by #4
   \ifnum\count\tempfor=0
    \advance\refnumber by 1  \count\tempfor=\refnumber \fi
 \testone=\count\tempone \testtwo=\count\temptwo \testthr=\count\tempthr
 \testfor=\count\tempfor
 \sortfour\testone\testtwo\testthr\testfor
   \test=\testthr \advance\test by -1
   \ifnum\testtwo=\test   \test=\testtwo \advance\test by -1
    \ifnum\testone=\test  \test=\testfor \advance\test by -3
     \ifnum\testone=\test \ [\the\testone--\the\testfor]
     \else \ [\the\testone--\the\testthr,\the\testfor]
     \fi
    \else  \test=\testfor \advance\test by -1
     \ifnum\testthr=\test \ [\the\testone,\the\testtwo--\the\testfor]
     \else\ [\the\testone,\the\testtwo,\the\testthr,\the\testfor]
     \fi
    \fi
   \else \ [\the\testone,\the\testtwo,\the\testthr,\the\testfor]
   \fi}

\def\reffive#1#2#3#4#5{\tempone=100 \advance\tempone by #1
   \ifnum\count\tempone=0
    \advance\refnumber by 1  \count\tempone=\refnumber \fi
    \temptwo=100 \advance\temptwo by #2
   \ifnum\count\temptwo=0
    \advance\refnumber by 1  \count\temptwo=\refnumber \fi
    \tempthr=100 \advance\tempthr by #3
   \ifnum\count\tempthr=0
    \advance\refnumber by 1  \count\tempthr=\refnumber \fi
    \tempfor=100 \advance\tempfor by #4
   \ifnum\count\tempfor=0
    \advance\refnumber by 1  \count\tempfor=\refnumber \fi
    \tempfiv=100 \advance\tempfiv by #5
   \ifnum\count\tempfiv=0
    \advance\refnumber by 1  \count\tempfiv=\refnumber \fi
 \testone=\count\tempone \testtwo=\count\temptwo \testthr=\count\tempthr
 \testfor=\count\tempfor \testfiv=\count\tempfiv
 \sortfive\testone\testtwo\testthr\testfor\testfiv
  \test=\testthr \advance\test by -1
  \ifnum\testtwo=\test   \test=\testtwo \advance\test by -1
   \ifnum\testone=\test  \test=\testfor \advance\test by -3
    \ifnum\testone=\test \test=\testfiv \advance\test by -4
     \ifnum\testone=\test\ [\the\testone--\the\testfiv]
     \else\ [\the\testone--\the\testfor,\the\testfiv]
     \fi
    \else \ [\the\testone--\the\testthr,\the\testfor,\the\testfiv]
    \fi
   \else  \test=\testfor \advance\test by -1
    \ifnum\testthr=\test \test=\testfiv \advance\test by -2
     \ifnum\testthr=\test \ [\the\testone,\the\testtwo--\the\testfiv]
     \else \ [\the\testone,\the\testtwo--\the\testfor,\the\testfiv]
     \fi
    \else\ [\the\testone,\the\testtwo,\the\testthr,\the\testfor,\the\testfiv]
    \fi
   \fi
  \else \test=\testfor \advance\test by -1
   \ifnum\testthr=\test \test=\testfiv \advance\test by -2
    \ifnum\testthr=\test\
[\the\testone,\the\testtwo,\the\testthr--\the\testfiv]
    \else\ [\the\testone,\the\testtwo,\the\testthr,\the\testfor,\the\testfiv]
    \fi
   \else\ [\the\testone,\the\testtwo,\the\testthr,\the\testfor,\the\testfiv]
   \fi
  \fi}

\def\refitem#1#2{\temp=#1 \advance \temp by 100 \setbox\count\temp=\hbox{#2}}

\def\sortfive#1#2#3#4#5{\sortfour#1#2#3#4\relax
   \ifnum#5<#4\relax \test=#5\relax #5=#4\relax
     \ifnum\test<#3\relax #4=#3\relax
       \ifnum\test<#2\relax #3=#2\relax
         \ifnum\test<#1\relax  #2=#1\relax  #1=\test
         \else #2=\test \fi
       \else #3=\test \fi
     \else #4=\test \fi \fi}

\def\sortfour#1#2#3#4{\sortthree#1#2#3\relax
    \ifnum#4<#3\relax \test=#4\relax #4=#3\relax
       \ifnum\test<#2\relax #3=#2\relax
          \ifnum\test<#1\relax #2=#1\relax #1=\test
          \else #2=\test \fi
       \else #3=\test \fi \fi}

\def\sortthree#1#2#3{\sorttwo#1#2\relax
       \ifnum#3<#2\relax \test=#3\relax #3=#2\relax
          \ifnum\test<#1\relax #2=#1\relax #1=\test
          \else #2=\test \fi \fi}

\def\sorttwo#1#2{\ifnum#2<#1\relax \test=#2\relax #2=#1\relax #1=\test \fi}


\def\setref#1{\temp=100 \advance\temp by #1
   \ifnum\count\temp=0
    \advance\refnumber by 1  \count\temp=\refnumber \fi}

\def\printreference{\totalnumber=\refnumber
           \advance\totalnumber by 1
           \itemnumber=0
           \loop \advance\itemnumber by 1  
                 \ifnum\itemnumber<\totalnumber
                   \item{\the\itemnumber .} \unhbox\itemnumber \repeat}

%
%
%
  \magnification=1200
  \hsize=6.0 truein
  \vsize=8.5truein
  \topskip=20pt            
  \fontdimen1\tenrm=0.0pt  
  \fontdimen2\tenrm=4.0pt  
  \fontdimen3\tenrm=7.0pt  
  \fontdimen4\tenrm=1.6pt  
  \fontdimen5\tenrm=4.3pt  
  \fontdimen6\tenrm=10.0pt 
  \fontdimen7\tenrm=2.0pt  
  \baselineskip=14.0pt plus 1.0pt minus 0.5pt  
  \lineskip=1pt plus 0pt minus 0pt             
  \lineskiplimit=1pt                           
  \parskip=2.5pt plus 5.0pt minus 0.5pt
  \parindent=15.0pt
%
%

 2
\font\bx=cmr8
\def\lsim{\; \raise0.3ex\hbox{$<$\kern-0.75em\raise-1.1ex\hbox{$\sim$}}\; }
\def\gsim{\; \raise0.3ex\hbox{$>$\kern-0.75em\raise-1.1ex\hbox{$\sim$}}\; }
\def\jump{\vskip 1truecm}
\def\jumpx{\vskip 1truecm\noindent}
\def\jumpp{\vskip 2truecm}
\def\jum{\vskip 0.5truecm}
\def\eV{\rm eV}

\def\MeV{\rm MeV}

\def\nue{\nu_{\rm e}}
\def\num{\nu_{\mu}}
\def\nut{\nu_{\tau}}

\def\1{\aa}
\def\vek#1{{\rm\bf #1}}

\def\del{\partial}
\def\ra{\rangle}
\def\la{\langle}

\def\nuebar{\bar{\nu}_{\rm e}}

%
%

\def\lr{\leftrightarrow}

%

\def\frac#1#2{{#1\over#2}}

%
%
\newcount\eqnumber
\eqnumber=1
\def\chaphead{}

\def\new{\hbox{(\chaphead\the\eqnumber}\global\advance\eqnumber by 1}
\def\eqref#1{\advance\eqnumber by -#1 (\chaphead\the\eqnumber
     \advance\eqnumber by #1 }
\def\first{\hbox{(\chaphead\the\eqnumber{a}}\global\advance\eqnumber by 1}
\def\last{\advance\eqnumber by -1 \hbox{(\chaphead\the\eqnumber}\advance
     \eqnumber by 1}
\def\eq#1{\advance\eqnumber by -#1 equation (\chaphead\the\eqnumber
     \advance\eqnumber by #1}
\def\eqnam#1{\xdef#1{\chaphead\the\eqnumber}}


\def\eqt#1{Eq.~({{#1}})}


\def\npr{neutron--to--proton\ ratio}
\def\He{\hbox{$^4He$}}
\def\SM{Standard\ Model}

\centerline {\bf ASPECTS OF NEUTRINO COSMOLOGY
\footnote{*}{\bx\baselineskip=9pt Invited talk at the XVI Kasimierz
Meeting on elementary Particle Physics, Poland, May 1993.}
}
\jumpp
\centerline{KARI ENQVIST}
\centerline{\sl Nordita, Blegdamsvej 17, DK-2100 Copenhagen \O , Denmark}
\jump
\centerline{ABSTRACT}
\jumpx
\vbox{\bx\baselineskip=12pt\parindent15pt\noindent\narrower{Primordial
nucleosynthesis constrains the properties of light, stable neutrinos.
Apart from the well--known limit on the number of neutrino species, there
are also bounds on neutrino masses and magnetic moments. I discuss also
sterile neutrinos and neutrino propagation in a primordial magnetic field,
such as could be the origin of the observed galactic magnetic fields.}
}
\jump\noindent
{\bf 1. Introduction}
\jum
The LEP results have made it clear that there are only three stable light
left--handed neutrinos which couple to Z.
Neutrinos might be exactly massless, but even if they have masses, we do not
know whether they should be Dirac or Majorana masses.
 For $\nue$
there exist several mass measurements\ref{3}, all yielding $m_{\nue}\lsim 10$
eV (although the average $\la m_{\nue}^2\ra$ is negative). The mass
limit on $\num$ has recently been revised upward\ref{3} to 500 keV,
although it is likely
to be improved soon.
The current limit on $\nut$ mass is $m_{\nu_\tau}\le 31$ MeV\ref{1}.
If neutrino masses are of the Majorana type, the non--observation of
neutrinoless double beta decay $(Z,A)\to (Z+2,A)+2e^-$ implies that\ref{4}
$\la m_\nu\ra\equiv\sum U_{ei}^2m_{\nu_i}\lsim 1.2$ eV; barring accidental
cancellations between the mixing matrix elements $ U_{ei}$, the Majorana
mass of $\nue$ should thus be less than about 1 eV.

If neutrinos are Dirac particles, in the \SM\ they will have small
one--loop induced magnetic moments, given by $\mu_\nu=3.1\times
10^{-19}(m_\nu/\eV)\mu_B$ where $\mu_B$ is the
Bohr magneton. Studies of $\nue e^-\to\nue e^-$ and $\num e^-\to\num e^-$
elastic scattering can be used to derive the  experimental
bounds\ref{18}   $\mu_{\nue}<
1.1\times 10^{-9}\mu_B$ and $\mu_{\num}<7.4\times 10^{-10}\mu_B$.
The best limit on $\mu_{\nut}$ is obtained from  D decays in   beam
dump experiments\ref{19}, yielding $\mu_{\nut}<5.4\times 10^{-7}\mu_B$.
Majorana neutrinos cannot have diagonal magnetic moments, but
if neutrinos mix, they will have non--zero transition magnetic moments.

There are a number of astrophysical constraints on neutrino
properties, mainly based on cooling of stars and the supernova SN1987A.
Energy would be transferred too fast from the inner core of the
supernova if left--handed neutrinos flip over to right--handed neutrinos.
These would then freely stream out\ref{21}, thus affecting the energetics of
the supernova. The most
recent numerical study\ref{22} employs a supernova code that includes
also neutrino--nucleon scattering
and has a higher meson density in the core compared with previous estimates.
Therefore the importance of processes like $\pi+N\to N+\nu\bar{\nu}$ is
enhanced, and one obtains a rather stringent upper limit of 3 keV on
the neutrino Dirac mass term.

There is also
the well--known cosmological limit on the sum of all stable neutrino masses,
whether Dirac or Majorana, which
can be obtained by requiring that the relic neutrino mass density
does not exceed the upper limit of the density of the universe.
One finds\eqnam{\summnu}
$$
\sum m_{\nu_i}=92\Omega_{0\nu}h^2\ \eV\; ,
\eqno\new)
$$
where $\Omega_{0\nu}$ is the energy density of neutrinos today, in units of
the critical density, and $h$ is the Hubble parameter in units of 100
kms$^{-1}$/Mpc. The age estimates of the universe imply\ref{5} that
$\Omega_0h^2\lsim 0.25$
so that $\sum m_{\nu_i}\lsim 23$ eV. If there exists a cosmological constant,
then the limit is somewhat relaxed; for $\rho_{\rm vac}=0.8\rho_{\rm crit}$
one obtains $\sum m_{\nu_i}\lsim 35$ eV.

For cosmology the difference between Dirac and Majorana neutrinos
is significant because that difference may be crucial for primordial
nucleosynthesis of light
elements\ref{20}. If neutrinos are Dirac particles, then the right handed
chirality state $\nu_R$, as well as the left--handed antineutrino
$\bar{\nu}_L$,
must exist. This means that at very high temperatures, for each
neutrino flavour, there were four spin--degrees of freedom
in equilibrium (in the \SM\ equilibration of right--handed neutrinos
takes place e.g. through the one--loop
induced magnetic moment). Later right--handed neutrino interactions, being
very weak, decoupled and their number densities were diluted by subsequent
annihilations. It is then essential that this decoupling occurs at high enough
temperature so that the relic density of right--handed neutrinos at the onset
of primordial nucleosynthesis is small enough. The current nucleosynthesis
limit on the maximum number of extra degrees of freedom, quantified in units of
two--component massless fermions, is often quoted as\ref{24} \eqnam{\nulimit}
$$\delta N_\nu\simeq 0.3
\eqno\new)
$$
This obviously sets a limit on the strength of interactions which turn
left--handed neutrinos into right--handed ones. These interactions will
necessarily involve spin--flip operators such as the mass and the
magnetic moment.

More generally, nucleosynthesis imposes a constraint on the equilibration rate
of any sterile neutrino, be it left-- or right--handed. This turns out to be
a very useful way to limit oscillations between sterile and active neutrino
species.

\jumpx\noindent
{\bf 2. Primordial nucleosynthesis}
\jum
At temperatures $T\gg {\cal O}(1)$ MeV neutrons and protons were kept in
chemical equilibrium through the weak processes $\nue n\lr pe,\ en\lr p\nuebar$
and
$n\lr pe\nuebar$. At that time the relative number of neutrons was simply given
by the equilibrium ratio\eqnam{\npratio}
$$X(T)\equiv{n_n(T)\over n_n(T)+n_p(T)}=\left (1+e^{\Delta m/T}\right)^{-1},
\eqno\new)
$$
where $\Delta m=m_n-m_p$. The \npr\ froze out at about $T\simeq 0.7$ MeV, after
which the neutron population was still reduced by free neutron decay until
$T\simeq 0.1$ MeV, at which point photons no longer were energetic enough
to prevent protons and neutrons to combine to form deuterium. Consequently
reactions like $D+D\to {^3He}+n,\ T+p;\ D+(p,T,{^3He})\to {^3He}+\gamma,\
\He+p(n)$
and ${^3He}+{^3He}\to \He+2p$ helped to build up about 25\%\ \He\ and traces of
$^3He,\ D$ and heavier elements such as lithium and beryllium. Therefore,
assuming the neutron life--time is known, the primordial abundance of
\He\ reflects essentially directly the \npr\
at the freeze--out, which depends only on the ratio $\Gamma_{n\lr p}(T)/H(T)$,
where $H(T)$ is the Hubble rate.

The theoretical Helium abundace thus depends mainly on two things: the
number of degrees of freedom at the freeze--out which appear in the
Hubble rate, given by $H=(8\pi^3g_*(T)/90)^{1/2}T^2/M_{Pl}$ where
$g_*(T)=\sum g_B(T)+{7\over 8}g_F(T)$ is the effective number of
degrees of freedom; and the number density of electron neutrinos, which
can affect the neutron--to--proton rate $\Gamma_{n\lr p}(T)$. In
addition, one needs to know the baryon density of
the universe, as well as the neutron life--time, which recently has been
revised\ref{20} to $10.26\pm 0.03$ min. There is also some inherent
uncertainty in the nuclear reaction rates, but for \He\ the
main uncertainty comes from neutron life--time. A fit to data
then yields\ref{24}\eqnam{\olive}
$$
N_\nu=3.0-0.8\ln\eta_{10}+19\left({Y_{\rm p}-0.228\over 0.228}\right )
-15\left({\tau_{\rm n}-889.8\over 889.8}\right ),
\eqno\new)
$$
where $Y_{\rm p}$ is the primordial \He-abundance, $\tau_{\rm n}$ is the
neutron life--time in seconds and $\eta_{10}\lsim 2.8$ is the baryon number
in units of $10^{-10}$.

\vbox to 12truecm{\bx\baselineskip=12pt\vfil\noindent Regression analyses
of He mass
fraction against O and N abundances, with  1$\sigma$ limits. Larger
and smaller circles represent higher and lower weights, respectively, open
circles are objects not enriched by Wolf-Rayet stars, and a
few typical error bars are shown (from ref.\reftext{28}).}
\jum
The primordial abundances of light elements can be deduced from
observations in various ways. For instance, one measures the relative
\He-abundance in extragalactic ionized
hydrogen regions with different metallicities. The result is
then interpolated to zero metallicity, which implies\ref{26}
a primordial abundance $Y_{\rm p}=0.228\pm 0.005$ (neglecting possible
systematic errors).
Adopting a
conservative upper bound $Y_{\rm p}\lsim 0.24$ in \eqt{\olive} gives then
rise to the limit \eqt{\nulimit}. Some quite recent observations of
certain very low metallicity
objects might however change the fit somewhat\ref{90}, and there
are also some theoretical issues like the nucleon mass corrections on the
reaction rates\ref{25}, which also alter the computed value of $Y_{\rm p}$
slightly.
It is nevertheless clear that in any case primordial nucleosynthesis
precludes the appearance of extra degrees of freedom at the level of
$\delta N_\nu\ll 1$.

The fit \eqt{\olive} can in fact be used to impose a simultaneous bound on
the neutrino number density and the number of extra degrees of freedom.
Let us denote the relative $\nue$--abundance by $n_{\nue}$ and $\delta
n_{\nue}\equiv n_{\nue}-1$. Then, if $g$ is the number of (fermionic) degrees
of freedom additional to the \SM , one can show that nucleosynthesis requires
that\ref{27}
\eqnam{\yleistys}
$$
g-4.6\delta n_{\nue}\le \delta N_\nu\; ,
\eqno\new)
$$
with $\delta N_\nu$ deduced from observations as in \eqt{\nulimit}
or \eqt{\olive}. Electron neutrino density may change either due to
oscillations
or heavy particle decays, which take place after the decoupling.
\jumpx\noindent
{\bf 3. Nucleosynthesis constraints on Dirac neutrinos}
\jum
The  production rate for the 'wrong helicity' neutrino $\nu_{+}$,
and hence their mass,
must be small enough  so that it decouples  already before the
QCD phase transition which occurs at temperatures somewhere
between 100 and 400 MeV. In that case they will not participate in the entropy
transfer from quark--gluon plasma to particles in equilibrium,
and consequently their number and energy densities will be diluted
below levels that are acceptable for
primordial nucleosynthesis. Assuming decoupling just above the QCD
phase transition,  at nucleosynthesis
there would then remain a right--handed neutrino energy density which would
be equivalent to about 0.1 neutrino species.

The 'wrong helicity' neutrino  production rate was first estimated by Fuller
and Malaney\ref{12}, who argued that a
Dirac neutrino with a lifetime exceeding the nucleosynthesis
time scale $(t\sim{1}\ {\rm{s}})$ should have a mass less than about
300 keV. A more  detailed study of the scattering processes involved has
recently been carried out in ref.\reftext{6}.

The starting point is that before the QCD phase transition but below, say,
$T\simeq 0.5$ GeV, the fermions present in the Universe at
significant number densities were the leptons
and u, d, s and c quarks. There are altogether 47
separate $2\rightarrow 2$ reactions
with {\it (i)} no 'wrong--helicity' neutrinos in the initial state
and {\it (ii)} with at least one
$\nu_+^{\mu}$ or $\nu_+^{\tau}$
in the final state, which all need to be taken into consideration.
The constraint {\it (i)} is imposed on because each wrong helicity
neutrino in the initial (final) state introduces an additional
small factor $m_\nu^2/|{\bf p}|^2$
($m_\nu^2/|{\bf p}^{\prime}|^2$) to the cross section.
Here $|{\bf p}|$ and $|{\bf p}^{\prime}|$ are the absolute
values of the centre--of--mass momenta of the incoming
and outgoing particles, respectively.
Hence processes with more than one 'wrong--helicity'
neutrino can be ignored as compared to prosesses with only
one 'wrong--helicity' neutrino. In addition, there are also quark and
lepton decays which can produce
$\nu_+$'s.

The thermally averaged scattering rate reads\ref{6}
\eqnam{\scatter}
$$
\Gamma_{+}^{sc}=
{1\over{n_{\nu_+}^{eq}(T)}}\sum_{(12\rightarrow{34})}
\int{d^3{p}_1\over{(2\pi)^3}}{d^3{p}_2\over{(2\pi)^3}}
f(E_1/T)f(E_2/T){\sigma}_{+}^{(12\rightarrow{34})}j(p_1,p_2),
\eqno\new)
$$
where $n_{\nu_+}^{eq}$ is the equilibrium number density of
$\nu_+$'s, $f(E_i/T)$
are the Fermi--Dirac distributions of the incoming
particles, and
$j(p_1,p_2)=\sqrt{(p_1\cdot p_2)^2-m_1^2m_2^2}/E_1E_2$ is a
flux--related factor.
In \eqt{\scatter} we have neglected the final state Pauli blocking,
which is an about 10\%\ effect\ref{6}.
Thermally averaged decay rate is simply given by
$$
\Gamma_{+}^d=
{1\over{n_{\nu_+}^{eq}(T)}}\sum_{(1\rightarrow{234})}
\Gamma_{+}^{(1\rightarrow{234})}
\int{d^3{p}_1\over{(2\pi)^3}}
f(E_1/T){{m_1}\over{E_1}},
\eqno\new)
$$
where the factor $m_1/E_1$ arises from the Lorentz boost
of the decay rate.

Computing all the relevant processes in this approximation, and requiring that
$\Gamma_+^{sc}+\Gamma_+^{d}<H$ at $T=T_{\rm QCD}$, one finds\ref{7} for
$T_{\rm{QCD}}\simeq{(100)200}$ MeV that
the mass limits are $m_{\nu_\tau}\lsim (1180)740$ keV and
$m_{\nu_\mu}\lsim{(720)480}$ keV.

A similar line of argument can be used to set a limit on the magnetic moment
of a light neutrino. The $e^+e^-$ annihilation cross section for right--handed
neutrinos in photon--mediated scattering has been estimated to be\ref{9}
$$\sigma\simeq  {\mu_\nu\alpha^2\pi\over 6m^2_{\rm e}}\left({1-4m^2_{\rm e}/s
\over 1-4m^2_\nu /s}\right )^{1/2}\left( 1+8{m^2_\nu\over s}\right)
\left( 1+2{m^2_{\rm e}\over s}\right).
\eqno\new)
$$
Demanding decoupling of the magnetic moment induced interactions prior to
QCD phase transition yields the limit\ref{11}
$$\mu_\nu\lsim 5.2\times 10^{-11}\mu_B\left({200\ \MeV\over T_{QCD}}\right ).
\eqno\new)
$$
More stringent limits\ref{15} will however be obtained from red giants and
SN1987A.
One may also put a nucleosynthesis limit on the neutrino charge radius\ref{51}:
$\langle r^2\rangle\lsim 10^{-32}{\rm cm}^2$.

\vbox to 10truecm{\vfil\noindent\bx\baselineskip=12pt Nucleosynthesis limits
on Dirac neutrino masses, as a function of QCD phase transition temperature.
The allowed region is below the curves. (From ref.\reftext{6}).}
\jum
The reasoning described above applies to neutrinos with a mass less than 1 MeV.
A heavy (tau) neutrino with a mass in the MeV region would
have a more pronounced effect on nucleosynthesis than a light neutrino,
because  during the synthesis of the light elements the energy density of
the 'right--helicity' states of a heavy neutrino
would be comparable to or higher than that of a massless neutrino.
This is because at that epoch these states have already decoupled
($T^{\nu_{-}}_{dec}\sim$ few MeV). This has been shown\ref{13} to lead to an
excluded region
$0.5\ {\rm MeV}\lsim{m}_{\nu_\tau}\lsim 30\ {\rm MeV}$ for the tau neutrino
mass, provided $\tau_{\nu_\tau}\gsim{10^3}$ s.
(If $1\ {\rm s}\lsim\tau_{\nu_\tau}\lsim 10^3$ s,
the upper bound is somewhat weakened.) This limit, with minor modifications,
applies also to Majorana neutrinos.

If $\nut$ has a very large magnetic moment,  it will be kept in equilibrium
by photon--mediated annihilations  which help to decrease the $\nut$ number
density. In this way $\nut$ could actually escape the nucleosynthesis
constraint above. Following the relevant Boltzmann equations one finds\ref{9}
in the mass range $5-35$ MeV the bound $\mu_{\nut}\lsim 10^{-8}$, with a
slight mass--dependence. The bound is essentially determined by the upper
limit on $\He$.

\jumpx\noindent
{\bf 4. Sterile neutrinos}
\jum
If neutrinos have a non--zero mass, they might also mix with each other
exactly like the quarks. In the early universe the mixing between flavour
states is not  expected to affect nucleosynthesis\ref{72}, because the
number densities of different flavour states are (to a high accuracy)
equal due to thermal equilibrium. If the flavour states  however mix with
a sterile neutrino $\nu_s$, then
many interesting effects arise. The initial number density of a sterile
species may be assumed to be diluted by e.g. QCD phase transition, but
oscillations may help to fill up the density back to its thermal level.
Moreover, oscillations may deplete the $\nue$ population after the
decoupling at $T\simeq 2.3$ MeV but above the n/p freeze--out at
$T\simeq 0.7$ MeV.
Both effects affect primordial nucleosynthesis, as is evident from
\eqt{\yleistys}.

The oscillation pattern between sterile and active neutrinos is in the
heat bath of the early universe modified by the forward scattering of
the active species. The average effective energy for $\nue$ at
$1\ \MeV\lsim T\ll 100\ \MeV$ is
given by\ref{53}\eqnam{\effpot}
$$
V_{\rm e}=\sqrt{2}G_Fn_{\gamma}(L-AT^2/m_W^2),
\eqno\new)
$$
where $A\simeq 55$ and $L$ contains terms proportional to lepton and
baryon asymmetries;
if they are initially small enough, they will be dynamically driven
to zero\ref{54}. Oscillation will not be effective if non--forward
collisions destroy the coherence, which means that oscillations
can start only at temperatures close to decoupling.

The oscillating system is conveniently
described by a $2\times 2$ one--body density matrix
$\rho_\nu={1\over 2}P_0(I+{\bf P}\cdot\sigma)$, and the equations
of motion are\ref{55}\eqnam{\tolkuton}
$$
\eqalign{
{d\vek P\over dt}=&\vek V\times\vek P+(1-P_z){d\ln P_0\over dt}\hat{\vek z}
-(D+{d\ln P_0\over dt})(P_x\hat{\vek x}+P_y\hat{\vek y}),\cr
{dn_{\nu_\alpha}\over dt}=&F_0[C_\alpha(1-n^2_{\nu_\alpha})-(n^2_{\nu_\beta}
-n^2_{\nu_\alpha})-(n^2_{\nu_\gamma}-n^2_{\nu_\alpha})],\cr
\vek V=&V_0({\rm sin}2\theta\hat{\vek x}-{\rm cos}2\theta\hat{\vek z})
\frac{\delta m^2}{\eV^2}\frac{\MeV}{T}-V_{\nu_\alpha}\hat{\vek z}
\left({T\over\MeV}\right)^5.\cr
}\eqno\new)
$$
Here $\alpha\ne\beta\ne\gamma\subset\{ e,\mu,\tau\}$, $C_e=2.31,
\ C_\mu=C_\tau=0.51$ and $F_0=2.65\times 10^{-2}(T/\MeV)^5{\rm s}^{-1}$.
$V_0$ and $V_{\nu_\alpha}$ are the effective energies, generalized
from \eqt{\effpot} to include also changes in $n_{\nue}$, $D$ the damping
rate and the equation of
motion for $P_0$ is simply $dP_0/dt=dn_{\nu_\alpha}/dt$. The number densities
are
normalized such that $n_{\nu_\alpha}=1$ corresponds to a single neutrino
degree of freedom in chemical equilibrium.
A complete description of the evolution of the oscillating system
\eqt{\tolkuton} requires numerical study which reveals\ref{55} that the
large--angle MSW solution for the solar neutrino problem\ref{56}  is, in the
case of $\nu_{\rm e}-\nu_s$ mixing, ruled out by nucleosynthesis.  Similarly,
the nucleosynthesis constraint rules out\ref{57} $\num-\nu_s$ oscillations
as an explanation for the atmospheric neutrino puzzle\ref{58}. (For
a recent
investigation of sterile--active oscillation in the early universe,
including a numerical nucleosynthesis code, see ref.\reftext{71}.)

Sterile neutrinos has also been proposed as a candidate
for the cold component in a cold and hot mixture of dark matter,
which has become popular in the
view of the COBE detection\ref{31} of the anisotropy in the microwave
background. It has been argued\ref{32} that sterile neutrino with a
Majorana mass in the range 0.1-1.0 keV, slightly mixed with an ordinary
neutrino, would provide
warm dark matter and structure formation with more power on small scales
than in hot dark matter scenarios.

Another interesting possibility, first pointed out by Madsen\ref{33}, is to
have a heavy fermion $\nu_H$ which decays into a light sterile fermion and a
light boson. These may then be assumed to have decoupled prior to QCD phase
transition. If $\nu_H$ decay while they are still relativistic, solving for
the relevant Boltzmann equations one finds that\ref{34} equilibration in
the final state will be preceded by decays into very cold ($p\ll T$) bosons.
The bose condensation is effective provided the decay rate is fast enough,
or equivalently, if the decay temperature is large enough: $T_d>\sqrt{m_Hm_B}$
where $m_B$ is the boson mass. This would then be a natural way to generate
a mixture of hot and cold dark matter. Numerical studies show\ref{34} that one
obtains about 40\%\ cold bosons, but this could still not be enough as
the favoured cold/hot ratio appears to be\ref{35} 70/30.
\jumpx\noindent
{\bf 5. Spin rotation in magnetic fields}
\jum
Right handed neutrinos may also be produced by scattering of $\nu_L$
off a primodial magnetic field, if such exists. This provides an
interesting connection with neutrino properties and primordial
nucleosynthesis on one hand, and with the observed galactic magnetic fields
on the other hand. The galactic magnetic
fields, which are of the order of few $\mu$G, are believed to have
arisen from a
very weak seed field through the so--called galactic dynamo mechanism\ref{41}.
The differential rotation and turbulent motion inside a galaxy amplifies
the seed field exponentially until a saturation point is reached. The field
is observed indirectly by measuring the syncrotron radiation of the electrons
which traverse the galactic field, assuming equipartition of magnetic and
particle energies (this assumption has however recently been subject to
some discussion\ref{61}). Observationally not much is known about the
seed field. Some weak bounds on it may be obtained by requiring that the
growth time must be longer than galactic rotation period. The
observation\ref{62}
of the magnetic field in a spiral galaxy with z=0.395 seems to indicate that
the dynamo was saturated already some time ago, implying a relatively large
seed field. Moreover, the magnetic field of the Milky Way changes its direction
by about 180$^o$ between the Sagittarius and Orion arms, which
has been argued\ref{63}
to be an indication for a large seed field of the order of about $10^{-11}$ G.
In computer simulations\ref{64} seed fields of the order of $10^{-18}$ seem to
be
able to produce the oberved field strength (but not the reversals, which
however has been observed only in the Milky Way).

It has been argued\ref{42} that a random seed field of the correct
size will indeed be produced by fluctuations in the Higgs field at the
electroweak phase transition. The point is that at the electroweak phase
transition all the physical quantities should be
uncorrelated over distances greater than the horizon distance. This means
in particular that the Higgs field cannot be gauge rotated to point into
same direction in group space in every horizon volume\ref{50}. Therefore
there will appear physical
Higgs field gradients from which one may construct an electromagnetic field
$F_{ij}=-i(V_i^\dagger V_j-V_iV_j^\dagger)$
with $V_i=2\sqrt{{\rm sin}\theta_W/g}\del_i\phi/|\phi|$. This is
a random field, frozen to
the charges in the primeval plasma, and for any Gaussian distribution one
finds\ref{42}\eqnam{\rms}
$$B_{rms}\equiv \sqrt{\langle B^2\rangle}=B_0\left ({R_0\over R(t)}\right)^2
{1\over \sqrt{N}},
\eqno\new)
$$
where $N$ is the number of correlation lengths and $B_0\simeq m_W^2\simeq
10^{24}$ G. This yields today and at the intergalactic distances of
100 kpc a root mean
square field of $4\times 10^{-19}$ G with $\langle B\rangle =0$.

The direct cosmological consequences of the random field \eqt{\rms} are
expected to be minor\ref{42}, except possibly for Dirac neutrino spin flip and
hence nucleosynthesis. For instance, the magnetic energy density at
the horizon scale
is much smaller than radiation energy density. At the onset of
nucleosynthesis,
and at the horizon scale, the magnitude of the field
is about $B_{rms}\simeq
1500$ G. This is well below the nucleosynthesis limit\ref{70}
$B\lsim 10^{11}$ G on primordial magnetic field,
which is based both on the effects on the expansion rate and on nuclear
reactions rates.

To tackle  neutrino helicity change
in an external slowly varying random magnetic field it is best to use a
kinetic equation\ref{46} for the Wigner neutrino
spin distribution function ${\bf S}({\bf p}, {\bf x},t)$. It is also useful
to consider the combination \eqnam{\viki}
$$\eqalign{
\tilde {H}_{\perp}(t)e^{\pm i\alpha (t)}&=\mu_{\nu}(B_x(t)+E_y(t) \pm
i(
B_y(t)-E_x(t))\cr
&={\mu_\nu\over \sqrt{2}}\int{d^3k\over (2\pi)^3}
[\hat{B}_x({\bf k})e^{i{\bf k}\cdot{\bf x}}
 \hat{B}_x^\dagger({\bf k})e^{-i{\bf k}\cdot{\bf x}}
\pm i(\hat{B}_y({\bf k})e^{i{\bf k}\cdot{\bf x}}
\hat{B}_y^\dagger({\bf k})e^{-i{\bf k}\cdot{\bf x}})],\cr
\alpha (t) &= \arctan \Bigl ( {B_y(t)-E_x(t)\over B_x(t)+E_y(t)}\Bigr
) ,\cr
}\eqno\new)
$$
corresponding to neutrino propagation along $z$-axis so that
$$
\tilde {H}_{\perp}(t) = \mu_{\nu}\sqrt {B_x^2(t)+
B_y^2(t) +E_x^2(t) +E_y^2(t)+2[E_y(t)B_x(t)-B_y(t)E_x(t)]}.
$$
The basic starting point for studying neutrino spin flip in random
magnetic field is then the
neutrino spin kinetic equation for the $z$--component of the neutrino
spin\eqnam{\pitka}
${\bf S}(t)=\int d^3p{\bf S}({\bf p},t)$, which takes the form\ref{46}
$$\eqalign{
{dS_z(t)\over dt}&+ 2\tilde {H}_{\perp}(t)e^{i(\alpha - Vt)}\int
\tilde
{H}_{\perp}(t)e^{-i(\alpha -Vt)}S_z(t)dt \cr
&+ 2\tilde {H}_{\perp}(t)e^{-i(\alpha - Vt)}\int \tilde
{H}_{\perp}(t)e^{i(\alpha - Vt)}S_z(t)dt \cr
&= i\Bigl (C_{-1}\tilde {H}_{\perp}(t)e^{-i(\alpha - Vt)} -
C_{+1}\tilde
{H}_{\perp}(t)e^{i(\alpha - Vt)}\Bigr ), \cr} \eqno\new)
$$
and which depends on the fluctuating magnetic field squared. The constants
$C_{\pm 1}$ are determined from boundary
conditions,
and $\tilde H_{\perp}$, $\alpha (t)$ are given by \eqt{\viki}.
$V=3.45\times 10^{-20} \left ( {T/ {\rm MeV}}\right )^5\ {\rm MeV}$ is the
effective neutrino energy as obtained from \eqt{\effpot} with $L=0$.

To make use of the complicated kinetic equation ({\pitka}) one has to average
over the fluctuations. This can be achieved by assuming isotropy and
using Wick rules for the various averages\ref{48}:
$$
\eqalign{
\langle\hat{B}_i^\dagger({\bf k})\hat{B}_j({\bf k'})\rangle=
&\ (2\pi)^3\delta^{(3)}({\bf k}-{\bf k'})
\langle B_i^\dagger B_j\rangle,\cr
\langle B_i^\dagger B_j\rangle_{\bf k}=
&\ (\delta_{ij}-k_ik_j/k^2)\langle {\bf B^2}\rangle_{\bf k}.
}
\eqno\new)
$$
 Many of the terms in
\eqt{\pitka} actually vanish by virtue of the averaging procedure.
After some manipulations, one finally finds that the kinetic
equation  reduces to the simple expression\ref{48} \eqnam{\kinet}
$$
{d^3S_z(t)\over dt^3} + \omega^2_0{dS_z(t)\over dt} = 0, \eqno\new)
$$
where the neutrino spin rotation frequency $\omega_0$ is given
by
$$
\omega_0 = \sqrt {(\dot {\alpha} - V)^2 + 4\tilde {H}^2_{\perp}}.
\eqno\new)
$$
Solving \eqt{\kinet} with the appropriate boundary conditions one
immediately obtains\eqnam{\PLR}
$$
P_{\nu_L\leftrightarrow \nu_R} = {1+S_z(t)\over 2}={4\tilde {H}_{\perp}^2
\over
\omega^2_0}\sin^2(\omega_0t/2), \eqno\new)
$$
where the frequency  now reads
$$
\omega_0 = \sqrt {V^2 + 8\langle \tilde {H}^2_{\perp}\rangle + {6\over 5}
L^{-2}}.
\eqno\new)
$$
Here $L$ is a measure of the magnetic field energy density
inhomogeneity, and it is determined by the ratio
$$
L^{-2} = {\int k^2\langle {\bf B}^2\rangle_kd^3k/(2\pi )^3\over
\langle {\bf B}^2\rangle_{{\bf x}=0}}. \eqno\new)
$$

A small fluctuation length of the magnetic field thus effectively damps the
neutrino to spin flip
probability. Unfortunately, it is not clear what a realistic coherence length
of the primordial field should be. Let us however assume that above the
QCD phase transition the field
is coherent over  length scales of the order of the  (electron) neutrino
scattering
length $L_W=(4.0G_F^2T^5)^{-1}\simeq 3\times 10^{-7}l_H$ at $T=200\
\MeV$ ($l_H$ is the horizon length). We may then average  $P_{L\to R}$
over   $L_W$ to obtain the production rate. One finds that
right--handed neutrinos  would be in equilibrium at nucleosynthesis
unless
$$
\Gamma_{\nu_L\to \nu_R}=\langle P_{L\to R}\rangle L_W^{-1}\lsim H(T=T_{\rm
QCD}).
\eqno\new)
$$
 In essence, the
spin content of the thermal neutrino bath is thus tested by ordinary weak
collisions. From \eqt{\PLR} one obtains the bound
$$
\mu_\nu B(T=200\ \MeV,L_W)\lsim 7\times 10^{2}\mu_B{\rm G}.
\eqno\new)
$$
Assuming the scale dependence of the coherent field is known, we can find
the magnitude of the galactic seed field at $T_{\rm QCD}$. With the scaling
\eqt{\rms} one would then obtain the limit $\mu_\nu\lsim 2\times
10^{-10}\mu_B$; with $B\sim 1/N$ the limit would be\ref{47}
$\mu_\nu\lsim {6.5\times 10^{-34}\mu_B{\rm G}/B_{seed}(T_{now})}\simeq
2\times 10^{-16}\mu_B.$
(A similar constraint applies also for transition magnetic moments). In
the \SM\ this argument can also be turned the other way round to yield a limit
on the magnetic field strength at QCD phase transition, provided galactic
dynamo is the explanation for the observed galactic magnetic
fields:\eqnam{\anvar}
$$
B(T=200\ \MeV)\lsim {10^{21}{\rm G}\over \sum (m_{\nu_i}/\eV)}.
\eqno\new)
$$
Again depending on the scaling of the magnetic field to intergalactic
distances, \eqt{\anvar} provides a limit  on the sum of all neutrino
masses\ref{49}.
\jumpx
{\bf References}
\jum
\def\yana#1#2#3{Astron.\ Astrophys.\ {\bf #2}\ {(19#1)}\ {#3}}
\def\yanas#1#2#3{Astron.\ Astrophys.\ Suppl.\ {\bf #2}\ {(19#1)}\ {#3}}
\def\yapj#1#2#3{Ap.\ J.\ {\bf #2}\ {(19#1)}\ {#3}}

\refitem{1}{H. Albrecht {\sl et al.}\ (ARGUS Collaboration), Phys. Lett. {\bf
B292} (1992) 221; D. Cinabro {\sl et al.}\ (CLEO Collaboration), Phys. Rev.
Lett. {\bf 70} (1993) 3700.}
\refitem{2}{220}
\refitem{3}{See e.g. R.G.H. Robertson in {\sl Proc. XXVI Int. Conf. on HEP} (J.
R. Stanford, ed.), AIP 1993, p. 140.}
\refitem{4}{A. Balysh {\sl et al.}, Phys. Lett. {\bf B283} (1992) 32.}
\refitem{5}{K. Freese and D. Schramm, Nucl. Phys. {\bf B233} (1989) 167.}
\refitem{6}{K.\ Enqvist, K.\ Kainulainen and V.\ Semikoz,  Nucl.\
 Phys.\ {\bf B374} (1992) 392.}
\refitem{7}{K.\ Enqvist and H.\ Uibo, Phys. Lett. {\bf B301} (1993) 376.}
\refitem{8}{E.L. Turner, Ap. J. {\bf 365} (1990) L43.}
\refitem{9}{L. Kawano, G.M. Fuller, R.A. Malaney and M.J. Savage, Phys. Lett.
{\bf B275} (1992) 487.}
\refitem{11}{M. Fukugita and S. Yazaki, Phys. Rev. {\bf D36} (1987) 3817.}
\refitem{15}{See e.g. G.\ Raffelt, Phys. Rep. {\bf 198} (1990) 1.}
\refitem{13}{E.W. Kolb {\sl et al.}, Phys.\ Rev. Lett.\ {\bf 67} (1991) 533;
A.D.\ Dolgov and I.Z.\ Rothstein, Phys.\ Rev.\ Lett.\ {\bf 71} (1993) 476.
}
\refitem{12}{G.M.\ Fuller and R.A.\ Malaney, Phys.\ Rev.\ {\bf D43}
(1991) 3136.}
\refitem{18}{D.A. Krakauer {\sl et al.}, Phys. Lett. {\bf B252} (1990) 177.}
\refitem{19}{A.M. Cooper-Sarkar {\sl et al.}, Phys. Lett. {\bf B280} (1992)
153.}
\refitem{41}{Ya.B. Zeldovich, A.A. Ruzmaikin and D.D. Sokoloff,
{\sl Magnetic Fields in Astrophysics} (McGraw-Hill, New York, 1980);
E.N. Parker, {\sl Cosmological Magnetic Fields} (Oxford Univ. Press,
Oxford, 1979); A.A. Ruzmaikin, A.A. Shukurov and D.D. Sokoloff,
{\sl Magnetic Fields of Galaxies} (Kluwer, Dordrecht, 1988).
}
\refitem{42}{K.\ Enqvist and P.\ Olesen, preprint NBI-HE-93-33.}
\refitem {44} {C. Aneziris and J. Schechter, Int. J. Mod. Phys. {\bf A6}
(1991) 2375.}
\refitem {45}{A.Yu. Smirnov, Sov. Phys. JETP Lett. {\bf 53} (1991)
291.}
\refitem {46}{V. Semikoz, preprint NORDITA 92/60, to appear in Phys. Rev. {\bf
D}.}
\refitem {47} {K. Enqvist, P. Olesen and V.B. Semikoz, Phys. Rev. Lett. {\bf
69} (1992) 2157.}
\refitem{48}{K.\ Enqvist and V.\ Semikoz, Phys. Lett. {\bf B312} (1993) 310.}

\refitem{49}{K. Enqvist, V. Semikoz, A. Shukurov and D. Sokoloff, preprint
NORDITA 92/91 AP, to appear in Phys. Rev. {\bf D}.}

\refitem{50}{T. Vachaspati, Phys. Lett. {\bf B265} (1991) 258.}
\refitem{51}{J. Grifols and E. Masso, Mod. Phys. Lett. {\bf A2} (1987) 205.}
\refitem{52}{G.F. Giudice, Phys. Lett. {\bf B251} (1990) 461; Mod. Phys.
Lett. {\bf A6} (1991) 851.}
\refitem{53}{D. Notz\"old and G. Raffelt, Nucl. Phys. {\bf B307} (1988) 924;
K. Enqvist, K. Kainulainen and J. Maalampi, Nucl. Phys. {\bf B349} (1991) 754.}
\refitem{54}{K. Enqvist, K. Kainulainen and J. Maalampi, Phys. Lett. {\bf B349}
(1990) 186.}
\refitem{55}{K. Enqvist, K. Kainulainen and M. Thomson, Nucl. Phys. {\bf B373}
(1992) 498.}
\refitem{56}{J.N. Bahcall, {\sl Neutrino Astrophysics}, Cambrigde University
Press (Cambridge 1989).}
\refitem{57}{K. Enqvist, K. Kainulainen and M. Thomson, Phys. Lett. {\bf B288}
(1992) 145.}
\refitem{58}{R. Becker-Szendy {\sl et al.}(IMB Collaboration), Phys. Rev. {\bf
D46} (1992) 3720; K.S. Hirata {\sl et al.}(Kamiokande II Collaboration), Phys.
Lett. {\bf B280} (1992) 146; for a discussion of implications on neutrino
oscillations, see W. Frati {\sl et al.}, Phys. Rev. {\bf D48} (1993) 1140.}
\refitem{20}{For a recent review on constraints imposed by primordial
nucleosynthesis, see R.A. Malaney and G.J. Mathews, Phys. Rep. {\bf 229} (1993)
145.}
\refitem{21}{R. Gandhi and A. Burrows, Phys. Lett. {\bf B} (1990) 19; {\it
ibid.} {\bf B261} (1991) 519; A. Burrows, R. Gandhi and M.S. Turner, Phys. Rev.
Lett. {\bf 68} (1992) 3834.}
\refitem{22}{R. Mayle, D.N. Schramm, M.S. Turner and J.R. Wilson, preprint
FERMILAB-PUB-92-381-A. }
\refitem{23}{V.V. Smith, D.L. Lambert and P.E. Nissen, Ap.J. in press}
\refitem{24}{K. Olive, D.N. Schramm, G. Steigman and T. Walker, Phys. Lett.
{\bf B236} (1990) 454; T. Walker et al., Ap. J. {\bf 376} (1991) 393.}
\refitem{25}{D. Seckel, preprint BA-93-16.}
\refitem{26}{B. Pagel, E. Simonson, R. Terlevich and M. Edmunds, Mon. Not. R.
Astron. Soc. {\bf 255} (1992) 325.}
\refitem{27}{K. Enqvist, K. Kainulainen and M. Thomson, Nucl. Phys. {\bf B373}
(1992) 498.}
\refitem{31}{G.F. Smoot {\sl et al.}, Ap. J.{\bf 396} (1992) L1.}
\refitem{32}{S. Dodelson and L. Widrow, preprint FERMILAB-PUB-93-057-A.}
\refitem{33}{J. Madsen, Phys. Rev. Lett. {\bf 69} (1992) 571.}
\refitem{34}{N. Kaiser, R.A. Malaney and G.D. Starkman, Phys. Rev. Lett. {\bf
71} (1993) 1128.}
\refitem{35}{M. Davis, F.J. Summers and D. Schlegel, Nature {\bf 359} (1992)
393; A.N. Taylor and M. Rowan-Robinson,  Nature {\bf 359} (1992) 396.}
\refitem{61}{X. Chi and A.W. Wolfendale, Nature {\bf 362} (1993) 610.}
\refitem{62}{P.P. Kronberg, J.J. Perry and E.L.H. Zukowski, Ap. J. {\bf 387}
(1992) 300.}
\refitem{63}{A. Poezd, A. Shukurov and D. Sokoloff, Mon.\ Not.\ R.\ Astron.\
Soc.
(submitted).}
\refitem{64}{J.S. Panesar and A.H.\ Nelson, \yana{92}{264}{77};
D.\ Elstner, R.\ Meinel, and R.\ Beck, \yanas{92}{94}{587};
D.\ Moss, A.\ Brandenburg, K.J.\  Donner and M.\ Thomasson,
\yapj{93}{409}{179};
R.\ Meinel, D.\  Elstner and G.\ R\"udiger, \yana{90}{236}{L33};
K.J.\ Donner and A.\ Brandenburg, \yana{90}{240}{289}.
}
\refitem{70}{B. Cheng {\sl et al.}, preprint FERMILAB-PUB-93-260-A.}
\refitem{71}{X. Shi, D.N. Schramm and B.D. Fields, preprint FERMILAB-PUB-93.}
\refitem{72}{P.G. Langacker {\sl et al.}, Nucl. Phys. {\bf B266} (1986) 669;
M.J. Savage {\sl et al.}, Ap. J. {\bf 368} (1991) 1.}
\refitem{90}{G. Steigman, private communication.}
\printreference
\end